\documentclass[12pt,a4paper, amsmath,amssymb]{article}
	\usepackage[margin=0.8in]{geometry}
	\usepackage{amssymb}
	\usepackage{amsmath}
	\usepackage{amsfonts}
	\usepackage[utf8]{inputenc}
	\usepackage{graphicx}
	\usepackage{float}
	\usepackage{xcolor}
	\usepackage{caption}
	\definecolor{bluemunsell}{rgb}{0.0, 0.5, 0.69}
	\usepackage{float}
	\usepackage{adjustbox}
	\usepackage{relsize}
	\usepackage[colorlinks,linkcolor = bluemunsell,
	urlcolor  = bluemunsell,
	citecolor = bluemunsell,
	anchorcolor = bluemunsell]{hyperref}\usepackage{hyperref}
	\usepackage[normalem]{ulem}

	\allowdisplaybreaks
	\usepackage{xcolor}
	\usepackage{makecell}
	\usepackage{multirow}
	\usepackage{tikz}

	\usepackage{cite}
	\usepackage{tikz}
	\usepackage{ctable}
	\usepackage{feynmp-auto}
	\definecolor{green}{rgb}{0.0, 0.56, 0.0}
	
	\providecommand{\openone}{\leavevmode\hbox{\small1\kern-3.8pt\normalsize1}}

	\definecolor{green}{rgb}{0.0, 0.56, 0.0}

	\usepackage{etoolbox}
	\makeatletter

	\providecommand{\openone}{\leavevmode\hbox{\small1\kern-3.8pt\normalsize1}}

	\makeatother
	
\begin{document}
	
	\title{
	}
	
	\begin{center}
	{\Large \textbf{Large Hadron Collider Signatures of Exotic Vector-Like \\ \vspace{0.25cm} Quarks within the 2-Higgs Doublet Model Type-II  
	}}
	\thispagestyle{empty}
	\def\thefootnote{\fnsymbol{footnote}}
	\vspace{1cm}
	
	{\sc
	A. Arhrib$^{1,2}$\footnote{\href{mailto:aarhrib@gmail.com}{aarhrib@gmail.com}},
	R. Benbrik$^3$\footnote{\href{mailto:r.benbrik@uca.ac.ma}{r.benbrik@uca.ac.ma}},
	M. Boukidi$^3$\footnote{\href{mailto:mohammed.boukidi@ced.uca.ma}{mohammed.boukidi@ced.uca.ma}},
	S. Moretti$^{4,5}$\footnote{\href{mailto:stefano.moretti@physics.uu.se}{stefano.moretti@physics.uu.se}; \href{mailto:s.moretti@soton.ac.uk}{s.moretti@soton.ac.uk}}\\}
	
	\vspace{1cm}
	{\sl\small
	$^1$Abdelmalek Essaadi University, Faculty of Sciences and Techniques, Tangier, Morocco\\
	\vspace{0.1cm}
	$^2$Department of Physics and CTC, National Tsing Hua University, Hsinchu 30013, Taiwan \\\vspace{0.1cm}
	
	$^3$Polydisciplinary Faculty, Laboratory of Fundamental and Applied Physics, Cadi Ayyad University, Sidi Bouzid, B.P. 4162, Safi, Morocco\\
	\vspace{0.1cm}
	$^4$Department of Physics \& Astronomy, Uppsala University, Box 516, SE-751 20 Uppsala, Sweden\\
	
	\vspace{0.1cm}
	$^5$School of Physics \& Astronomy, University of Southampton, Southampton, SO17 1BJ, United Kingdom
	}
	\end{center}
	\begin{abstract}
	\noindent
	We study the decay of the exotic Vector-Like Quarks (VLQs) $X$ and $Y$, with $5/3$ and $-4/3$ units of electric charge, respectively, within the 2-Higgs Doublet Model (2HDM). 
	Building on our previous studies of Vector-Like Top and Bottom  (VLT and VLB) quarks, we now investigate the characteristics of $X$ and $Y$ in the alignment limit of a Type-II Yukawa structure and show that, in the framework of such a 2HDM, one can have large  non-Standard Model (SM) decay rates  of the $X$ and$ Y$ states. Our analysis focuses on their potential detection at the Large Hadron Collider (LHC), based on their pair production followed by a variety of  both SM and non-SM   decay patterns. In order to distinguish between doublet and triplet representations of the VLQs $X$ and $Y$, we uncover specific signatures that can provide insights into this particular architecture of Beyond the SM (BSM) physics.
	\end{abstract}
	
	\def\thefootnote{\arabic{footnote}}
	\setcounter{page}{0}
	\setcounter{footnote}{0}
	
	\newpage
	
	\section{Introduction}
	\noindent 
	The discovery of the Higgs boson during LHC Run 1 at CERN provided an affirmation of the SM as a successful Electro-Weak (EW) theory. Most of the Higgs observables   as  analysed by the  ATLAS and CMS collaborations agree  well with SM predictions \cite{ATLAS:2012yve, CMS:2012qbp}. Even in presence of this success, though, there are a lot of signs pointing to the existence of BSM physics to explain dark matter, the gauge hierarchy, neutrino masses, etc., thereby necessitating further exploration into potential extensions of the SM, such as those  incorporating  VLQs.  Such  particles appear in various new physics scenarios  such as little Higgs constructs  \cite{Arkani-Hamed:2002ikv,Han:2003wu}, 
	composite Higgs models \cite{Agashe:2004rs,Bellazzini:2014yua}, extra dimensions \cite{Agashe:2006wa} and  new top-flavour models \cite{He:1999vp,Wang:2013jwa}.
	
	A plausible BSM framework  is provided by the 2-Higgs Doublet Model (2HDM) \cite{Branco:2011iw, Draper:2020tyq}, which introduces, besides the SM-like Higgs boson ($h$), extra Higgs states ($H$, $A$ and $H^\pm$). Within this framework, we previously studied VLT and VLB quarks \cite{Arhrib:2024tzm,Arhrib:2024dou}, by highlighting their implications for LHC phenomenology. Expanding upon such previous results, the current work analyses $X$ and $Y$ states, i.e., exotic VLQs,  specifically, in the alignment limit of the 2HDM with a 
	Type-II Yukawa structure.
	
	The inclusion of VLQs in the 2HDM framework modifies the phenomenology of the latter in a fundamental way, not only by expanding the particle spectrum but also by introducing new VLQ decay channels involving the extra Higgs bosons. These channels, in particular, include accessible decays into charged Higgs bosons ($H^\pm$), which can be rather light, owing to mixing effects and cancellations occurring in flavour observables between the SM top-quark contributions and the VLT ones \cite{Benbrik:2022kpo}, in turn constituting an hallmark signature of this BSM scenario.  In fact, also the exotic VLQs, $X$ and $Y$, offer a chance to test uniquely 
	our BSM scenario. Both $X$ and $Y$ VLQs can be copiously pair produced at the LHC  via gluon-gluon fusion similarly to  $t\bar{t}$ production, through Quantum Chromo-Dynamics  (QCD) interactions, i.e, via $pp\to X\bar X$ and $Y\bar Y$.  They can also be singly produced through either QCD or  EW interactions such as $pp\to q g \to W^* t^* \to Xtq'$, $pp\to q g \to W^* b\to Yb q$ \cite{Aguilar-Saavedra:2013qpa,Aguilar-Saavedra:2009xmz} or  $pp \to bq \to Yq'$ (via $t$-channel W exchange) \cite{Shang:2024wwy}. 
	In the SM with VLQs, the $X$ state would mostly decay into a $W$ boson and a top quark  whereas the $Y$ state would exclusively decay into a $W$ boson and bottom quark.  Therefore, the experimental signature for $X(Y)$ would be $2W^+2W^- b\bar{b}$($W^+W^-b\bar{b}$) \cite{CMS:2018ubm,CMS:2017mrm}. In particular, 
	we highlight here the similarities between the $Y$ search (assuming their pair production) and the one for VLTs,  which decay predominantly into $Wb$: in this case, the 
	limits on such a VLT  would also apply to the $Y$ state, as they would both produce $W^+W^- b\bar b $ events \cite{ATLAS:2018dyh}. Furthermore, note that the presence of  two bottom quarks in the $X$ search  results in significant jet activity as well as leptons with the same sign or else a single lepton plus light-quark jets, signatures that are heavily constrained \cite{CMS:2018ubm}. (We will account for limits emerging from all these processes in our forthcoming analysis.)
	
	In this study, we focus on thoroughly analysing the $X$ and $Y$  decay patterns, with particular attention on differentiating between doublet and triplet representations of VLQs.   
	In doing so, we will offer important new perspectives on the underlying structure of BSM physics, especially concerning the EW Precision Observables (EWPOs).  
	Additionally, this study also seeks to complement and extend our previous analyses of VLQs within the 2HDM  Type-II  \cite{Arhrib:2024tzm, Arhrib:2024dou},
	thereby adding to a more thorough understanding of their role in BSM scenarios.
	
	Despite the emphasis on the exotic VLQs, $X$ and $Y$, this study employs the same methodology as our earlier studies on VLTs and VLBs \cite{Arhrib:2024tzm, Arhrib:2024dou}. 
	Just like for the case of VLTs and VLBs therein, 
	we will demonstrate herein the importance of the non-SM decay channels at the LHC for the case of $X$ and $Y$, thereby adding some new results to the extensive literature on 2HDMs with VLQs  \cite{Benbrik:2015fyz, Arhrib:2016rlj, Aguilar-Saavedra:2013qpa, Badziak:2015zez, Angelescu:2015uiz, Aguilar-Saavedra:2009xmz, DeSimone:2012fs, Kanemura:2015mxa, Lavoura:1992np, Chen:2017hak, Carvalho:2018jkq, Moretti:2016gkr, Prager:2017hnt, Prager:2017owg, Moretti:2017qby, Deandrea:2017rqp, Aguilar-Saavedra:2017giu, Alves:2023ufm, Dermisek:2019vkc, Dermisek:2020gbr, Dermisek:2021zjd, Benbrik:2022kpo, Vignaroli:2012sf, Vignaroli:2015ama, Vignaroli:2012si}.
	
	The structure of this paper is as follows. Section 2 provides a comprehensive description of the theoretical model employed here. Section 3 presents and discusses the results of our analysis, focusing on the decay patterns of the  $X$ and $Y$ states. Finally, Section 4 summarises our findings and presents our conclusions.
	
	\section{Model description}
	
	\subsection{Formalism}
	This study builds upon the framework established in Refs. \cite{Arhrib:2024tzm, Arhrib:2024dou}, where the phenomenology of a 2HDM Type-II was investigated with the inclusion of VLQs. This work expands the analysis to the case of $X$ and $Y$ VLQs, while prior work concentrated on the decay dynamics of VLT and VLB quarks.

	Two Higgs doublet fields, $\Phi_1$ and $\Phi_2$, are present in the 2HDM. At the tree level, Flavour Changing Neutral Currents (FCNCs) would result if both Higgs doublets coupled to all fermions as in the SM. A $\mathbb{Z}_2$ symmetry is therefore enforced on the Higgs doublets, $\Phi_1 \to \Phi_1$ and $\Phi_2 \to -\Phi_2$ \cite{Glashow:1976nt}, in order to suppress such FCNCs at  tree level. 
	The resulting Higgs potential, with the $\mathbb{Z}_2$ symmetry softly broken by dimension-2 terms proportional to $m^2_{12}$, is expressed as:
	\begin{eqnarray} \label{pot}
	\mathcal{V} &=& m^2_{11}\Phi_1^\dagger\Phi_1+m^2_{22}\Phi_2^\dagger\Phi_2
	-\left(m^2_{12}\Phi_1^\dagger\Phi_2+{\rm h.c.}\right)
	\nonumber \\
	&&+\frac{1}{2}\lambda_1\left(\Phi_1^\dagger\Phi_1\right)^2
	+\frac{1}{2}\lambda_2\left(\Phi_2^\dagger\Phi_2\right)^2  +\lambda_3\Phi_1^\dagger\Phi_1\Phi_2^\dagger\Phi_2
	\nonumber \\
	&&+\lambda_4\Phi_1^\dagger\Phi_2\Phi_2^\dagger\Phi_1
	+\left[\frac{1}{2}\lambda_5\left(\Phi_1^\dagger\Phi_2\right)^2+{\rm h.c.}\right].
	\end{eqnarray}
	Choosing real Vacuum Expectation Values (VEVs) for the two Higgs doublet fields, $v_1$ and $v_2$, and demanding
	$m_{12}^2$ and $\lambda_5$ to be real, the potential is  CP-conserving.
	After EW Symmetry Breaking (EWSB) takes place, the spectrum of the 2HDM contains:  two CP-even Higgs bosons ($h$ and $H$, with $m_h<m_H$), one CP-odd Higgs boson ($A$) and a pair of charged Higgs bosons ($H^\pm$). Here, $h$ is identified as the observed SM-like Higgs particle observed at the LHC with $m_h=125$ GeV.
	The  independent parameters are here taken to be the four masses, $m_h$, $m_H$, $m_A$ and $m_{H^\pm}$,
	the soft breaking parameter $m_{12}$, the VEV ratio $\tan \beta = v_2/v_1$ and the mixing term $\sin(\beta-\alpha)$,
	where the angle $\alpha$ diagonalises the CP-even mass matrix. When we impose that no (significant) tree-level FCNCs are present in the theory using the (softly broken) $\mathbb{Z}_2$ symmetry, we end up with four different Yukawa versions of the model \cite{Branco:2011iw}.   Type-II is the Yukawa texture  where $\Phi_2$ couples to up-type quarks and $\Phi_1$ couples to 
	charged leptons and down-type quarks.

	The gauge invariant structures that have multiplets with definite ${SU}(3)_C \times {SU}(2)_L \times {U}(1)_Y$ quantum numbers appear in the interactions of the  VLQs with the SM states via renormalisable couplings. The set of VLQ representations is indicated by:
	\begin{align}
	& T_{L,R}^0  \,  \ , \,  B_{L,R}^0   & \text{(singlets)} \,, \notag \\
	& (X\,T^0)_{L,R} \,, \quad (T^0\,B^0)_{L,R}   \,, \quad (B^0\,Y)_{L,R}  & \text{(doublets)} \,, \notag \\
	& (X\,T^0\,B^0)_{L,R} \,, \quad (T^0\,B^0\,Y)_{L,R}  & \text{(triplets)} \,.
	\end{align}
	We use in this section a zero superscript to distinguish the weak eigenstates from the mass eigenstates. The electric charges of the  VLQs are $ Q_T = 2/3 $, $ Q_B = -1/3 $, $ Q_X = 5/3 $ and $ Q_Y = -4/3 $. (Note that then $T$ and $B$ carry the same electric charge as the SM top and bottom quarks, respectively.)
	
	The physical up-type quark mass eigenstates may, in general, contain non-zero $Q_{L,R}^0$ (with $Q$ being the VLQ field) components, when new fields $T_{L,R}^0$ of charge $2/3$ and non-standard isospin assignments are added to the SM. This situation leads to a deviation in their couplings to the $Z$ boson. Atomic parity violation experiments and the measurement of $R_c$ at LEP impose constraints on these deviations for the up and charm quarks which are  significantly stronger than those for the top quark.
	In the Higgs basis, the Yukawa Lagrangian contains the following terms:
	\begin{equation}
	-\mathcal{L} \,\, \supset  \,\, y^u \bar{Q}^0_L \tilde{H}_2 u^0_R +  y^d \bar{Q}^0_L H_1 d^0_R + M^0_u \bar{u}^0_L u^0_R  + M^0_d \bar{d}^0_L d^0_R + \rm {h.c}.
	\end{equation}
	Here, $u_R$ actually runs over $(u_R, c_R, t_R, T_R)$ and $d_R$ actually runs over $(d_R, s_R, b_R, B_R)$. 
	
	We now turn to the mixing of the new partners with the third generation, $y_u$ and $y_d$, which are $3\times 4$ Yukawa matrices. In fact, in the light of the above constraints, 
	it is very reasonable to assume that only the top quark $t$  ``mixes'' with $T$.
	In this case, the $2 \times 2$ unitary matrices $U_{L,R}^u$ define the relation between the charge $2/3$ weak and mass eigenstates:
	\begin{eqnarray}
	\left(\! \begin{array}{c} t_{L,R} \\ T_{L,R} \end{array} \!\right) &=&
	U_{L,R}^u \left(\! \begin{array}{c} t^0_{L,R} \\ T^0_{L,R} \end{array} \!\right)
	\nonumber \\
	&=& \left(\! \begin{array}{cc} \cos \theta_{L,R}^u & -\sin \theta_{L,R}^u e^{i \phi_u} \\ \sin \theta_{L,R}^u e^{-i \phi_u} & \cos \theta_{L,R}^u \end{array}
	\!\right)
	\left(\! \begin{array}{c} t^0_{L,R} \\ T^0_{L,R} \end{array} \!\right) \,.
	\label{ec:mixu}
	\end{eqnarray}
	In contrast to the up-type quark sector, the addition of new fields $B_{L,R}^0$ of charge $-1/3$ in the down-type quark sector results in four mass eigenstates $d,s,b,B$.
	Measurements of $R_b$ at LEP set constraints on the $b$ mixing with the new fields that are stronger than for mixing with the lighter quarks $d,s$. 
	
	In this case, then,  $2 \times 2$ unitary matrices $U_{L,R}^d$ define the dominant $b-B$ mixing as 
	\begin{eqnarray}
	\left(\! \begin{array}{c} b_{L,R} \\ B_{L,R} \end{array} \!\right)
	&=& U_{L,R}^d \left(\! \begin{array}{c} b^0_{L,R} \\ B^0_{L,R} \end{array} \!\right)
	\nonumber \\
	&=& \left(\! \begin{array}{cc} \cos \theta_{L,R}^d & -\sin \theta_{L,R}^d e^{i \phi_d} \\ \sin \theta_{L,R}^d e^{-i \phi_d} & \cos \theta_{L,R}^d \end{array}
	\!\right)
	\left(\! \begin{array}{c} b^0_{L,R} \\ B^0_{L,R} \end{array} \!\right) \,.
	\label{ec:mixd}
	\end{eqnarray}
	
	(More details on this Lagrangian formalism are shown in the Appendix.) 
	To ease the notation, we have dropped the superscripts $u$ ($d$) whenever the mixing occurs only in the up (down)-type quark sector. 
	Additionally, we sometime use the shorthand notations $s_{L,R}^{u,d} \equiv \sin \theta_{L,R}^{u,d}$, $c_{L,R}^{u,d} \equiv \cos \theta_{L,R}^{u,d}$, etc.
	
	This Lagrangian contains all the phenomenological relevant information:
	\begin{itemize}
	\item[(i)] the modifications of the SM couplings that might show indirect effects of new quarks can be found in the terms that do not contain heavy quark fields;
	\item[(ii)] the terms relevant for LHC phenomenology (i.e., heavy quark production and decay) are those involving a heavy and a light quark;
	\item[(iii)] terms with two heavy quarks are relevant for their contribution to oblique corrections.
	\end{itemize}
	In the weak eigenstate basis, the diagonalisation of the mass matrices makes the Lagrangian of the third generation and heavy quark mass terms such as
	
	\begin{eqnarray}
	\mathcal{L}_\text{mass} & = & - \left(\! \begin{array}{cc} \bar t_L^0 & \bar T_L^0 \end{array} \!\right)
	\left(\! \begin{array}{cc} y_{33}^u \frac{v}{\sqrt 2} & y_{34}^u \frac{v}{\sqrt 2} \\ y_{43}^u \frac{v}{\sqrt 2} & M^0 \end{array} \!\right)
	\left(\! \begin{array}{c} t^0_R \\ T^0_R \end{array}
	\!\right) \notag \\
	& & - \left(\! \begin{array}{cc} \bar b_L^0 & \bar B_L^0 \end{array} \!\right)
	\left(\! \begin{array}{cc} y_{33}^d \frac{v}{\sqrt 2} & y_{34}^d \frac{v}{\sqrt 2} \\ y_{43}^d \frac{v}{\sqrt 2} & M^0 \end{array} \!\right)
	\left(\! \begin{array}{c} b^0_R \\ B^0_R \end{array}
	\!\right) +\text{h.c.},
	\label{ec:Lmass}
	\end{eqnarray}
	with $M^0$ a bare mass
	term\footnote{As pointed out in the introduction, this bare mass term is not related to the Higgs mechanism. It is gauge-invariant and can appear as such in the Lagrangian, or it can be generated by a Yukawa coupling to a scalar multiplet that acquires a VEV $v' \gg v$.}, $y_{ij}^q$, $q=u,d$, Yukawa couplings and  $v=246$ GeV the Higgs VEV in the SM. Using the standard techniques of diagonalisation, the mixing matrices are obtained by
	\begin{equation}
	U_L^q \, \mathcal{M}^q \, (U_R^q)^\dagger = \mathcal{M}^q_\text{diag} \,,
	\label{ec:diag}
	\end{equation}
	with $\mathcal{M}^q$ the two mass matrices in Eq.~(\ref{ec:Lmass}) and $\mathcal{M}^q_\text{diag}$ the diagonals ones. 
	To check the consistency of our calculation,  we have verified that the corresponding $2 \times 2$ mass matrix reduces to the SM quark mass term if either of  the $T$ or $B$ quarks are absent.

	Notice also that, in multiplets with both $T$ and $B$ quarks, the bare mass term is the same for the up-and down-type quark sectors. For singlets and triplets one has $y_{43}^q = 0$ whereas for doublets $y_{34}^q=0$. Moreover, for the $XTB$ triplet one has $y_{34}^d = \sqrt 2 y_{34}^u$ and for the $TBY$ triplet one has $y_{34}^u = \sqrt 2 y_{34}^d$\footnote{We write the triplets in the spherical basis, hence, the $\sqrt 2$ factors stem from the relation between the Cartesian and spherical coordinates of an irreducible tensor operator of rank 1 (vector).}.
	
	The mixing angles in the left- and right-handed sectors are not independent parameters. From the mass matrix bi-unitary diagonalisation in Eq.~(\ref{ec:diag}) one finds:
	\begin{eqnarray}
	\tan 2 \theta_L^q & = & \frac{\sqrt{2} |y_{34}^q| v M^0}{(M^0)^2-|y_{33}^q|^2 v^2/2 - |y_{34}^q|^2 v^2/2} \quad \text{(singlets, triplets)} \,, \notag \\
	\tan 2 \theta_R^q & = & \ \frac{\sqrt{2}  |y_{43}^q| v M^0}{(M^0)^2-|y_{33}^q|^2 v^2/2 - |y_{43}^q|^2 v^2/2} \quad \text{(doublets)} \,,
	\label{ec:angle1}
	\end{eqnarray}
	
	with the relations:
	\begin{eqnarray}
	\tan \theta_R^q & = & \frac{m_q}{m_Q} \tan \theta_L^q \quad \text{(singlets, triplets)} \,, \notag \\
	\tan \theta_L^q & = & \frac{m_q}{m_Q} \tan \theta_R^q \quad \text{(doublets)} \,,
	\label{ec:rel-angle1}
	\end{eqnarray}
	with $(q,m_q,m_Q) = (u,m_t,m_T)$ and $(d,m_b,m_B)$, so one of the mixing angles is always dominant, especially in the down-type quark sector. In addition, for the triplets, the relations between the off-diagonal Yukawa couplings lead to relations between the mixing angles in the up-and down-type quark  sectors,
	\begin{eqnarray}
	\sin 2\theta_L^d & = & \sqrt{2} \, \frac{m_T^2-m_t^2}{m_B^2-m_b^2} \sin 2 \theta_L^u \quad \quad (X\,T\,B) \,, \notag \\
	\sin 2\theta_L^d & = & \frac{1}{\sqrt{2}} \frac{m_T^2-m_t^2}{m_B^2-m_b^2} \sin 2 \theta_L^u \quad \quad (T\,B\,Y) \,.
	\label{TBY-mix}
	\end{eqnarray}

	Due to non-zero mixing with SM quarks, the masses of the heavy VLQs depart from $M^0$. In the case of doublets and triplets, there is a relationship between the masses of the various multiplet components. Together, these relations demonstrate that a mixing angle, a heavy quark mass and a CP-violating phase that enters some couplings may be used to parametrise all multiplets except the $TB$ doublet: in fact, the latter is disregarded for the observables examined in this study. 
	
	From here on, we refer to such a model as the 2HDM+VLQs scenario and we will consider the  doublet and triplet representations, which offer the aforementioned exotic states.  
	Specifically, we focus on the $(XT)$ and $(BY)$ doublets as well as the $(XTB)$ and $(TBY)$ triplets, all within the alignment limit of the 2HDM, where $m_h = 125$ GeV is fixed and $m^2_{12}$ is chosen as $m_A^2\frac{\tan^2\beta}{1+\tan^2\beta}$.
	
	\subsection{Model Implementation and Validation}
	
	We now go into detail about how our BSM scenario is implemented.  We employed \texttt{2HDMC-1.8.0} \cite{Eriksson:2009ws} publich code 
	as the core framework for our 2HDM+VLQ setup.\footnote{A public release of this implementation is forthcoming. The analytical expressions for the Feynman rules governing the interaction vertices of the 2HDM+VLQ model have been integrated as a new class. Additionally, several new tree-level VLQ decay processes, including those involving Higgs bosons decaying into VLQs, have been explicitly coded.} To produce a correct mass spectrum and couplings, the Lagrangian components were first implemented in \texttt{FeynRules-2.3} \cite{Degrande:2011ua}.  With this configuration, we produced Universal FeynRules Output (UFO) interfaces for \texttt{MadGraph-3.4.2} \cite{Alwall:2014hca} as well as model files for \texttt{FeynArts-3.11} \cite{Hahn:2000kx,Kublbeck:1990xc} and \texttt{FormCalc-9.10} \cite{Hahn:2001rv,Hahn:1998yk}.
	We then confirmed the cancellation of Ultra-Violet (UV) divergences and verified the independence from the renormalisation scale across a few pertinent one-loop-level processes to make sure the approach was consistent.
	
	\subsection{Constraints}
	\label{sec-A}
	
	In this section, we outline the constraints applied to obtain our final results. 
	
	From a theoretical perspective, we imposed the following conditions.
	\begin{itemize}
	\item \textbf{Unitarity} constraints: The $S$-wave components of various (pseudo)scalar-(pseudo)scalar, (pseudo)scalar-gauge boson, and gauge-gauge boson scatterings must remain unitary at high energies~\cite{Kanemura:1993hm}.
	\item \textbf{Perturbativity} constraints: The quartic couplings of the scalar potential must satisfy $|\lambda_i| < 8\pi$ for $i = 1, \dots, 5$~\cite{Branco:2011iw}.
	\item \textbf{Vacuum stability} constraints: The scalar potential must be bounded from below and positive in any direction within the field space. Consequently, the $\lambda_i$ parameters must satisfy the conditions~\cite{Barroso:2013awa,Deshpande:1977rw}:
	\begin{align}
	\lambda_1 > 0,\quad\lambda_2>0, \quad\lambda_3>-\sqrt{\lambda_1\lambda_2} ,\nonumber\\ \lambda_3+\lambda_4-|\lambda_5|>-\sqrt{\lambda_1\lambda_2}.
	\end{align} 
	\item \textbf{EWPO limits}: The oblique parameters $S$ and $T$~\cite{Grimus:2007if} were employed to ensure that any parameter point in our model satisfies the following $\chi^2$ criteria within the 95\% Confidence Level (CL), i.e.,  aligned with global fit results \cite{Molewski:2021ogs}:
	\begin{align}
	&S= 0.05 \pm 0.08,\quad T = 0.09 \pm 0.07,\nonumber \\&  \rho_{S,T} = 0.92 \pm 0.11\hspace{0.5cm}(\text{for~~}U=0). 
	\end{align}
	A detailed discussion on EWPO contributions in VLQ scenarios can be found in \cite{Arhrib:2024tzm, Benbrik:2022kpo,Abouabid:2023mbu}. Notably, the unitarity, perturbativity, vacuum stability as well as $S$ and $T$ constraints were enforced using \texttt{2HDMC-1.8.0} \cite{Eriksson:2009ws}.
	\end{itemize}
	
	On the experimental front, we considered the following constraints.
	\begin{itemize}
	\item \textbf{SM-like Higgs boson properties}: These were evaluated using \texttt{HiggsSignal-3} \cite{Bechtle:2020pkv,Bechtle:2020uwn} via \texttt{HiggsTools} \cite{Bahl:2022igd}, requiring that relevant quantities (such as signal strengths) satisfy $\Delta\chi^2 = \chi^2 - \chi^2_{\mathrm{min}}$ within 95\% CL ($\Delta\chi^2 \leq 6.18$) across 159 observables.
	\item \textbf{Direct search constraints}: Constraints from collider searches at LEP, Tevatron and LHC were considered at 95\% CL, utilizing \texttt{HiggsBounds-6} \cite{Bechtle:2008jh,Bechtle:2011sb,Bechtle:2013wla,Bechtle:2015pma} via \texttt{HiggsTools}, including the latest searches for neutral and charged scalars.
	\item \textbf{$b \to s\gamma$ constraints}: To comply with $b \to s\gamma$ limits, the charged Higgs boson mass was set at 600 GeV\footnote{The analysis in Ref. \cite{Benbrik:2022kpo} suggested that incorporating VLQs into the 2HDM Type-II could relax this limit through large mixing angles and cancellations in flavour processes, EWPOs impose constraints that tend to keep the charged Higgs mass near  the standard 2HDM Type-II limit, hence, there is a need of some fine-tuning to lower the latter, which we decided not to enforce here.}.
	\item \textbf{LHC direct search constraints for VLQs}: The LHC direct search constraints are critical for setting exclusion limits on VLQs. Current LHC searches primarily focus on the SM decay modes of VLQs, specifically $X/Y \to Wt/b$, where these channels dominate with Branching Ratios (${\cal BR}$s) of 100\%. However, in our scenarios, where new decay modes involving charged Higgs bosons ($X/Y \to H^\pm t/b$) are introduced, the existing LHC limits must be applied more carefully. Specifically, the SM-based constraints are only directly applicable when the ${\cal BR}$s for $X/Y \to Wt/b$ remain 100\%, and contributions from the exotic decays $X/Y \to H^\pm t/b$ are negligible.
	
	To incorporate this into our analysis, we applied the existing ATLAS and CMS limits on both single and pair production of VLQs, using the exclusion criterion ${\sigma_{\mathrm{theo}}}/{\sigma_{\mathrm{obs}}^{\mathrm{LHC}}} < 1$ to retain only the parameter points that satisfy these experimental bounds.
	
	Our exclusion results are summarised as follows.
	
	\begin{itemize}
	\item \textbf{For VLQ $\boldsymbol{Y}$:} ATLAS constraints on single production \cite{ATLAS:2018dyh, ATLAS:2016ovj, ATLAS:2016scx} exclude mixing angles larger than approximately 0.2 for masses below 1.4 TeV. Pair production limits from ATLAS \cite{ATLAS:2024gyc, ATLAS:2017nap} exclude masses below 1.7 TeV, assuming $\mathcal{BR}(Y \to Wb) = 100\%$. CMS provides similar exclusions for both single and pair production \cite{CMS:2017fpk, CMS:2017ynm}.
	
	\item \textbf{For VLQ $\boldsymbol{X}$:}  ATLAS single production exclusion limits are reported in \cite{ATLAS:2018alq}, while pair production constraints exclude masses below 1.47 TeV, assuming $\mathcal{BR}(X \to Wt) = 100\%$ \cite{ATLAS:2018alq, ATLAS:2017nap, ATLAS:2022tla, ATLAS:2016sno, ATLAS:2018mpo, ATLAS:2015vzd, ATLAS:2015uaw}. CMS imposes comparable exclusions for both single and pair production \cite{CMS:2018dcw, CMS:2018ubm}.
	\end{itemize}
	
	Fig.~\ref{fig_LHC} presents an example of the exclusion limits for VLQ $Y$ in the $(m_Y, s^d_R)$ plane, considering both the 2HDM+$BY$ doublet (left) and 2HDM+$TBY$ triplet (right) scenarios. The theoretical results are superimposed with the 95\% C.L. limits from ATLAS \cite{ATLAS:2018dyh}. This figure illustrates how the combined theoretical and experimental constraints define the allowed parameter space, particularly when considering both SM and non-SM decays.
	
	\begin{figure}[H] \centering \includegraphics[width=0.45\textwidth,height=0.45\textwidth]{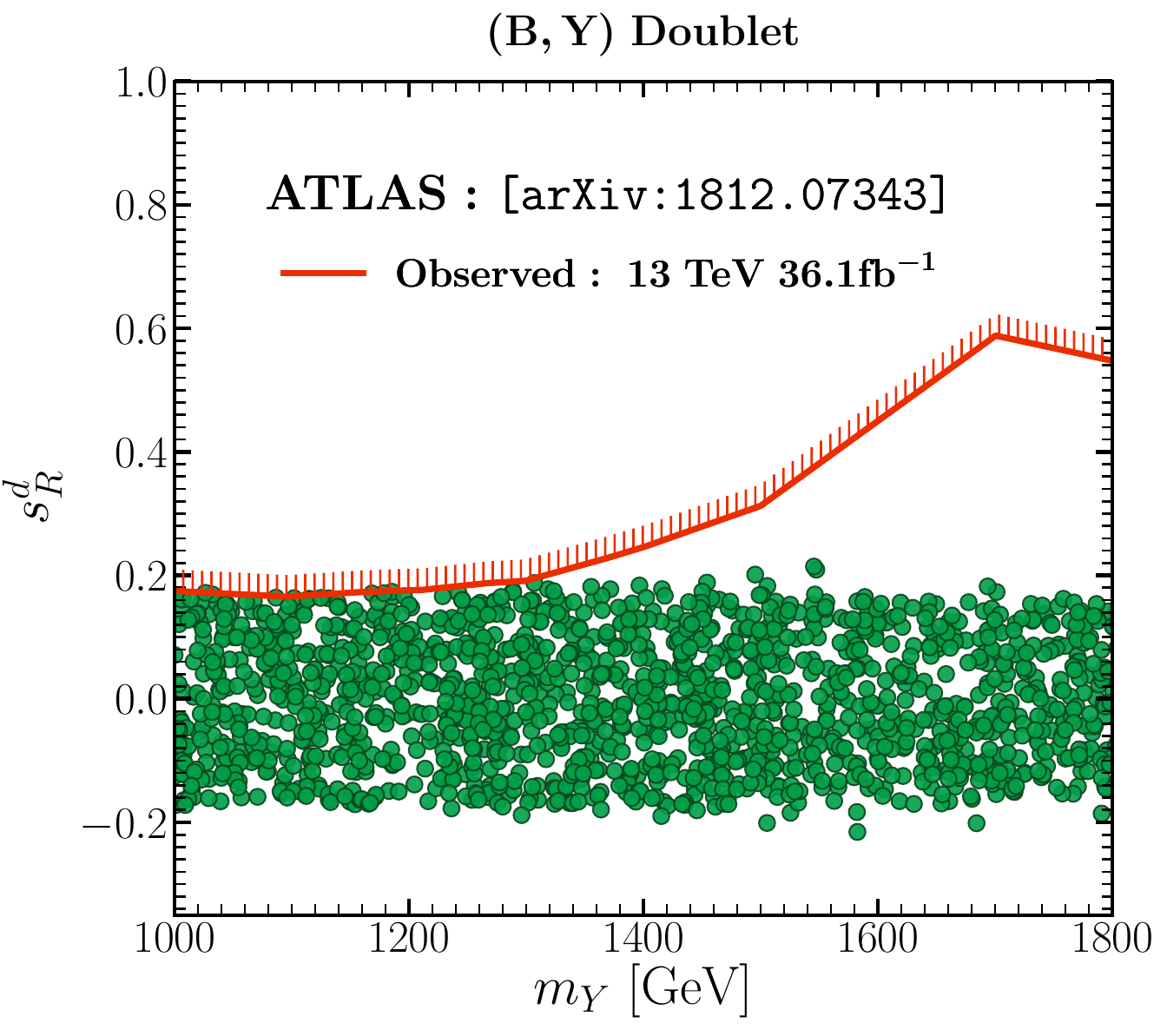} \includegraphics[width=0.45\textwidth,height=0.45\textwidth]{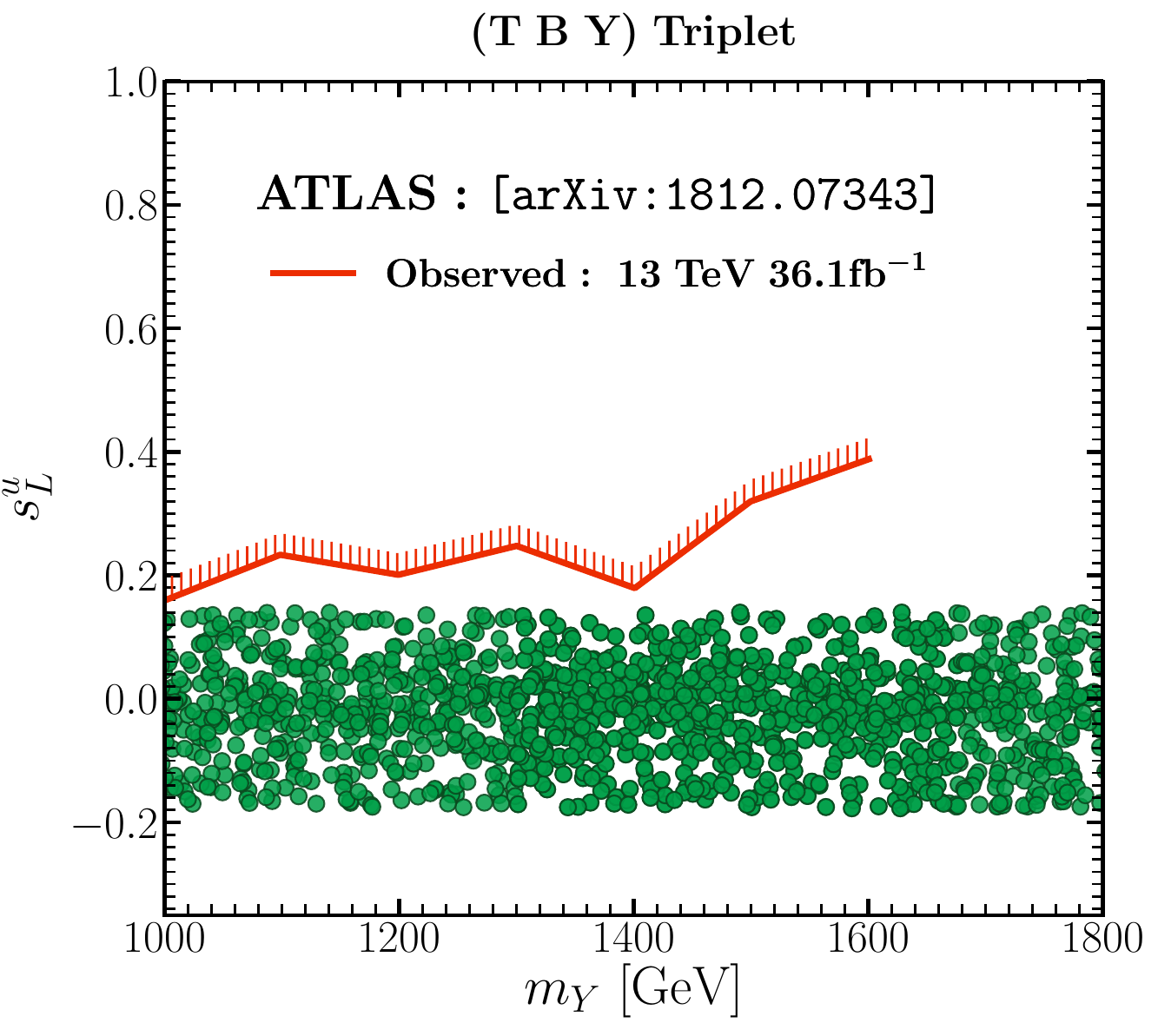} \caption{Allowed points following the discussed theoretical and experimental constraints in the ($m_Y, s^d_R$) plane for the 2HDM+$BY$ doublet scenario (left) and 2HDM+$TBY$ triplet scenario (right), superimposed onto the ATLAS \cite{ATLAS:2018dyh} 95\% C.L. observed upper limits on the couplings $s^d_R$ and $s^u_L$, respectively.} \label{fig_LHC} \end{figure}

	\end{itemize}
	\section{Numerical results}
	In this section, we present our findings on the various VLQ representations involving $X$ and $Y$ within the 2HDM Type-II framework, in relation to cross sections and ${\cal BR}$s.
	
	\subsection{2HDM with the $(XT)$ Doublet}
	
	For the scenario where the SM is extended with an $(XT)$ doublet, the properties of the VLQ $X$ are primarily determined by the mixing angle $\theta_R^t$ and the new top quark mass $m_T$. The mixing $\theta_L^t$ can be computed using Eq.~(\ref{ec:rel-angle1}) once $\theta_R^t$ is specified.
	The mass of the VLQ with an exotic electric charge of $5/3$ is given by \cite{Aguilar-Saavedra:2013qpa}:
	
	\begin{equation}
	m_X^2 = m_T^2\cos\theta_R^2 + m_t^2\sin\theta_R^2.\label{eq_XT}
	\end{equation}
	
	This mass relation is independent of the parameters associated with the 2HDM Higgs sector at tree level, although the latter influences the model viability when subjected to EWPOs constraints.

\begin{figure}[H]
	\centering
	\includegraphics[width=0.85\textwidth,height=0.4\textwidth]{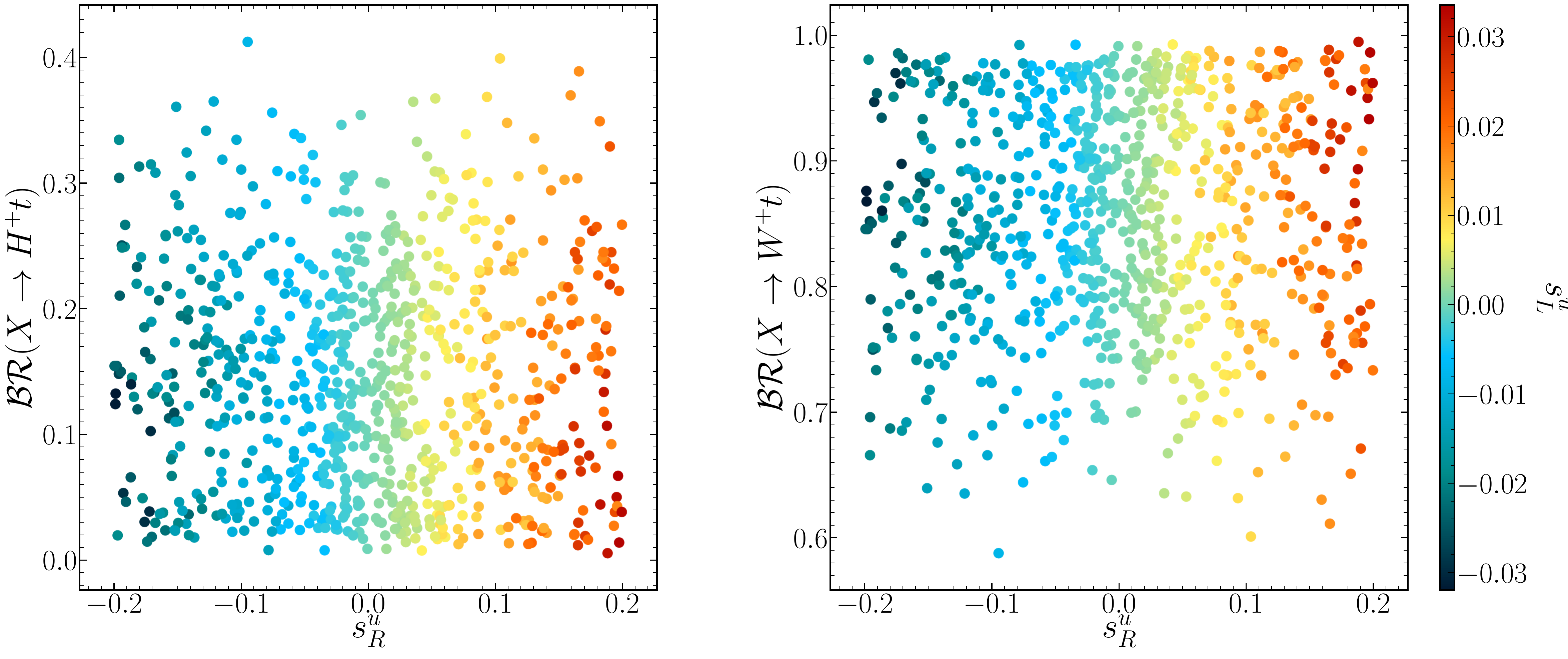}
	\caption{Scatter plots of parameter points that satisfy all imposed constraints in the $\mathcal{BR}(X\to H^+t)$ versus $s_R^u$ (left) and $\mathcal{BR}(X\to W^+t)$ versus $s_R^u$ (right) planes, with the colour bar representing $s_L^d$.}
	\label{fig1}
	\end{figure}

	\begin{table*}[t!]
	\centering
	{\setlength{\tabcolsep}{1.25cm}
	\begin{tabular}{cc}
		\toprule\toprule
		Parameters  & Scanned ranges \\
		\toprule
		$m_h$   & $125.09$ \\
		$m_A$  & [$400$, $800$] \\
		$m_H$  & [$400$, $800$] \\
		$m_{H^\pm}$  & [$600$, $800$] \\
		$\tan\beta$ & [$1$, $20$] \\
		$m_{X,Y}$   & [$1000$, $2000$] \\    
		$s_L^{u,d}$  & [$-0.5$, $0.5$] \\
		$s_R^{u,d}$  & [$-0.5$, $0.5$] \\
		\toprule\toprule
	\end{tabular}}
	\caption{2HDM and VLQ parameters for all scenarios with their scanned ranges. Masses are in GeV. 
	}
	\label{table1}
	\end{table*}
		Following the parameter scan detailed in Tab.~\ref{table1}, we investigated the $\mathcal{BR}$s for the VLQ $X$ decays into $H^+t$ and $W^+t$ as functions of $s_R^u$, as illustrated in Fig.~\ref{fig1}. The colour bar reflects variations in $s_L^u$. The $\mathcal{BR}$ for $X \to H^+t$ is limited to a maximum of 41\%, primarily due to the absence of the right-handed coupling $Z_R^{Xt}$. Additionally, the left-handed coupling $Z_L^{Xt}$, which is proportional to $s_R^u$, is constrained by EWPOs to remain small ($ |s_R^u|\le0.22$). Any further increase in this decay mode is limited by the exclusion of larger $\tan\beta$ values from LHC searches for BSM Higgs bosons, particularly in the $H^+ \to t\bar{b}$ channel \cite{ATLAS:2021upq}. In contrast, the SM decay channel $X \to W^+t$ consistently achieves a $\mathcal{BR}$ of 100\% across different values of $s_R^u$ and $s_L^u$.
	
	\begin{figure}[H]
	\centering
	\includegraphics[width=0.85\textwidth,height=0.4\textwidth]{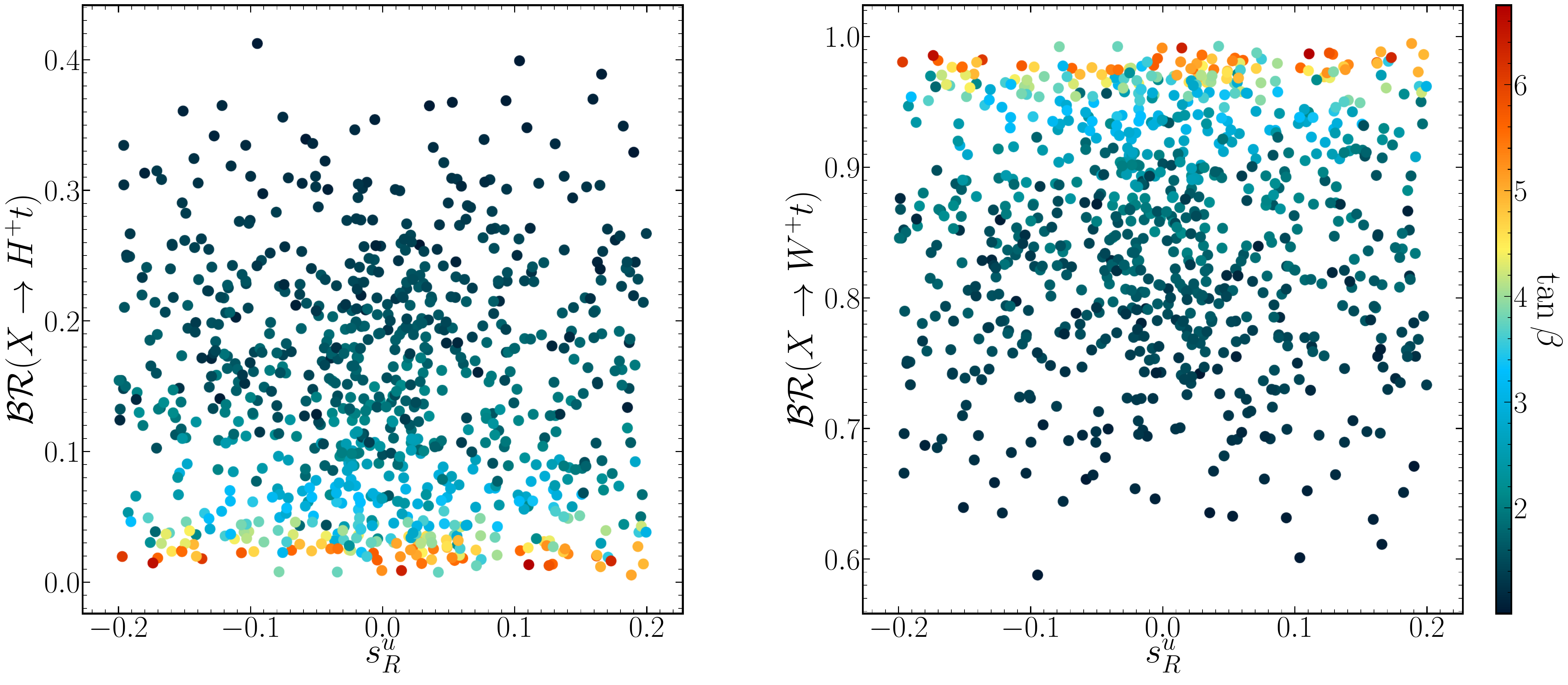}
	\caption{Similar to Fig.~\ref{fig1}, but with $\tan\beta$ shown in the colour bar.}
	\label{fig2}
	\end{figure}

	In Fig.~\ref{fig2}, we present the same data as in Fig.~\ref{fig1}, but now colour coded according to $\tan\beta$. The results show that $\mathcal{BR}(X\to H^+t)$ reaches its maximum at low $\tan\beta$, while the $W^+t$ decay dominates at intermediate values of $\tan\beta$\footnote{ It is important to note that large values of $\tan\beta$ are excluded by LHC searches for BSM Higgs bosons, specifically in the $H\to\tau\tau$ channel \cite{ATLAS:2020zms}.}.
	
	\subsection{2HDM with the $(BY)$ Doublet}
	
	In the case of the SM extended with a $(BY)$ doublet, the VLQ $Y$ is characterised by the mixing angle $\theta_R^b$ and the new bottom quark mass $m_B$. For a given $\theta_R^b$, the angle $\theta_L^b$ can be determined using Eq.~(\ref{ec:rel-angle1}). The mass of the VLQ with an exotic electric charge of $-4/3$, is given by \cite{Aguilar-Saavedra:2013qpa}:
	
	\begin{equation}
	m_Y^2 = m_B^2\cos\theta_R^2 + m_b^2\sin\theta_R^2.\label{eq_BY}
	\end{equation}
	
	Similar to the $(XT)$ case, this mass relation is independent of the 2HDM Higgs sector parameters at tree level, but the EWPO data constrains the overall viability of this BSM scenario.
	
\begin{figure}[H]
	\centering
	\includegraphics[width=0.85\textwidth,height=0.4\textwidth]{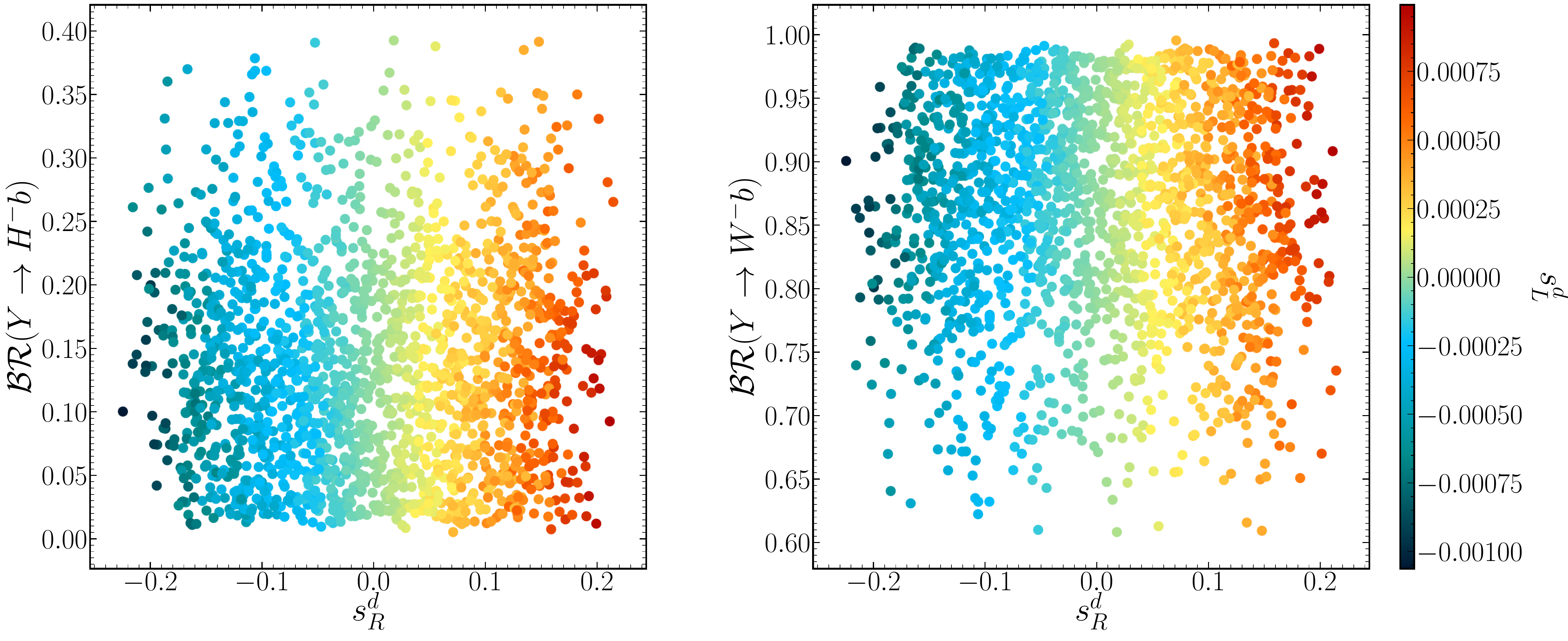}
	\caption{Scatter plots of parameter points that satisfy all imposed constraints in the $\mathcal{BR}(Y\to H^-b)$ versus $s_L^u$ (left) and $\mathcal{BR}(Y\to W^-b)$ versus $s_L^u$ (right) planes, with the colour bar indicating $s_L^d$.}
	\label{fig3}
	\end{figure}

		In Fig.~\ref{fig3}, we display the $\mathcal{BR}$s for the VLQ $Y$ decays into $H^-b$ and $W^-b$ as functions of $s_R^d$, with the colour bar indicating $s_L^d$. The results show that, analogous to the VLQ-$Y$ in the $(BY)$ doublet, the production of charged Higgs bosons from $Y$ can reach a maximum $\mathcal{BR}$ of 39\%, while the SM decay channel ($W^-b$) can achieve a 99\% $\mathcal{BR}$ for different values of $s_R^d$ and $s_L^d$.
	
	\begin{figure}[H]
	\centering
	\includegraphics[width=0.85\textwidth,height=0.4\textwidth]{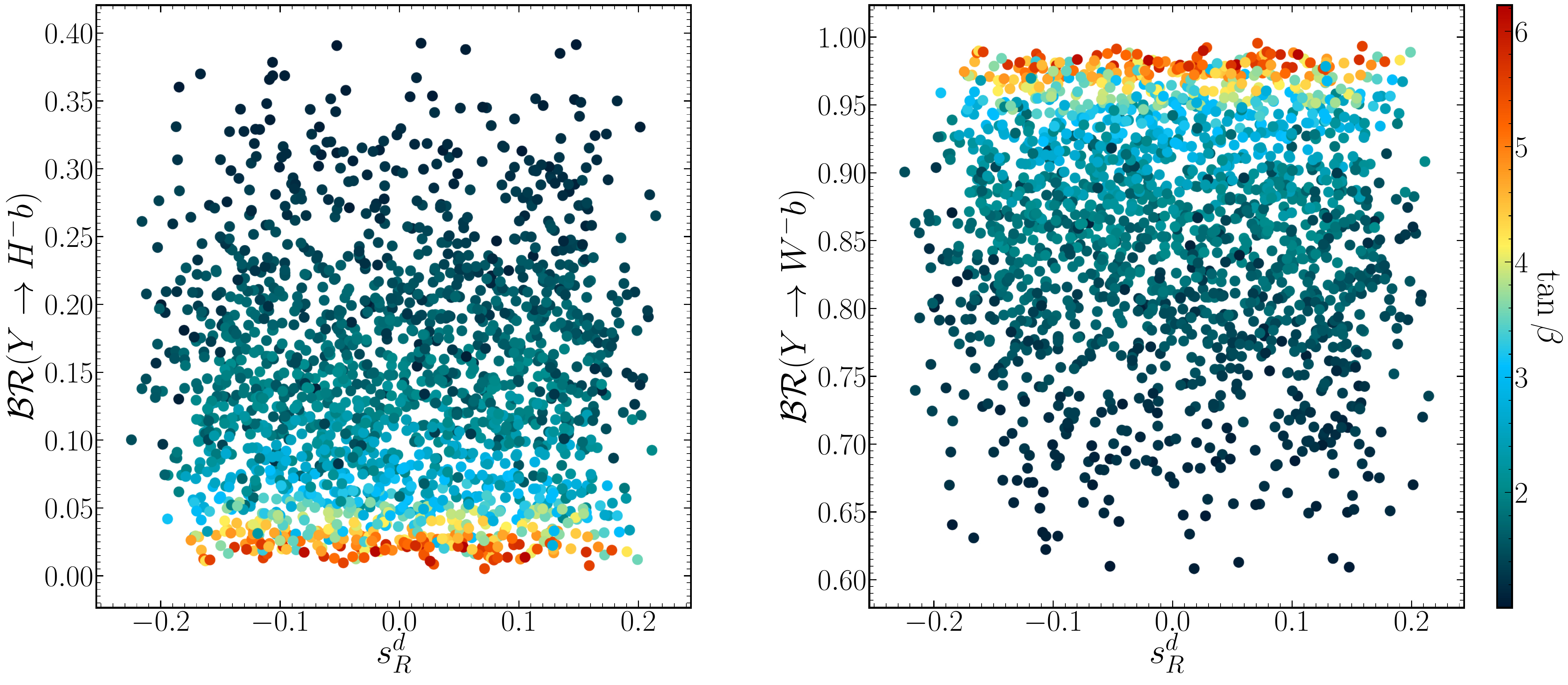}
	\caption{Similar to Fig.~\ref{fig3}, but with $\tan\beta$ shown in the colour bar.}
	\label{fig4}
	\end{figure}

	Finally, Fig.~\ref{fig4} shows the same samples as in Fig.~\ref{fig3}, but with $\tan\beta$ in the colour bar. Here, we observe that $\mathcal{BR}(Y\to H^-b)$ reaches its maximum at low $\tan\beta$, while the $W^-b$ decay is dominant at intermediate values of $\tan\beta$.
	\subsection{2HDM with the $(XTB)$ Triplet}
	
	In this subsection, we explore the $(XTB)$ triplet scenario. Before presenting the numerical results, it is essential to outline the parameterisation used in this model. The latter is determined by specifying the new top quark mass and one mixing angle, $\theta_L^t$, with the other parameters being derivable. Specifically, $\theta_R^t$ is computed using Eq.~(\ref{ec:rel-angle1}), while the mass of the $X$ quark is given by \cite{Aguilar-Saavedra:2013qpa}:
	
	\begin{eqnarray}
	m_X^2 &=& m_T^2 \cos^2\theta^u_L + m_t^2 \sin^2\theta^u_L\nonumber\\& =& m_B^2 \cos^2\theta_L^b + m_b^2 \sin^2\theta_L^b.
	\end{eqnarray}
	
	Utilising this relation between $m_T$ and $m_X$, along with the mixing relationships for up- and down-type quarks given in Eq.~(\ref{TBY-mix}), the mass of the new bottom quark $m_B$ can be derived as:
	
	\begin{eqnarray}
	m_B^2 = \frac{1}{2}\sin^2(2\theta^u_L)\frac{(m_T^2 - m_t^2)^2}{(m_X^2 - m_b^2)} + m_X^2.
	\end{eqnarray}
	
	The down-type quark mixing angle, $\theta_L^d$, is then obtained through:
	
	\begin{eqnarray}
	\sin(2\theta_L^d) = \sqrt{2}\frac{m_T^2 - m_t^2}{m_B^2 - m_b^2}\sin(2\theta^u_L).
	\end{eqnarray}
	
\begin{figure}[H]
	\centering
	\includegraphics[width=0.85\textwidth,height=0.4\textwidth]{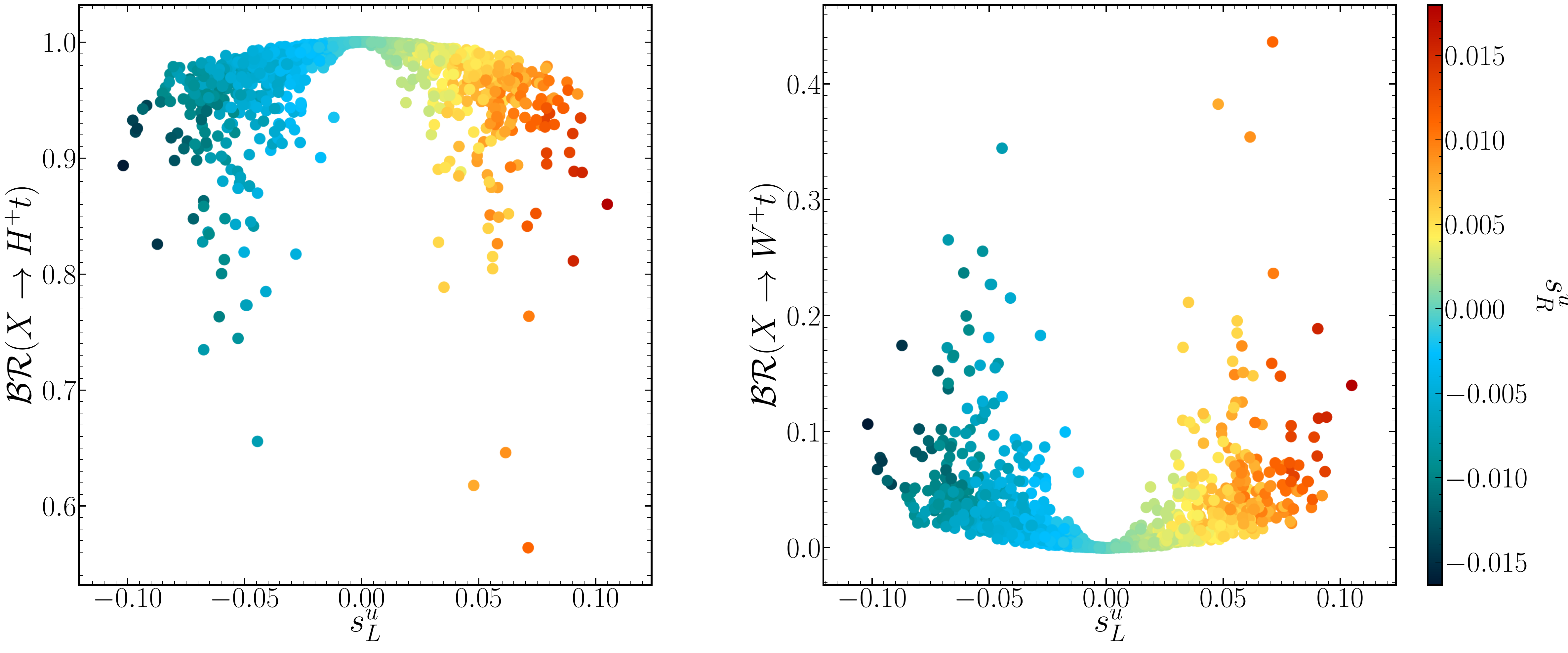}
	\caption{Scatter plots of parameter points that satisfy all imposed constraints in the $\mathcal{BR}(X\to H^+t)$ versus $s_L^u$ (left) and $\mathcal{BR}(X\to W^+t)$ versus $s_L^u$ (right) planes, with the colour bar representing $s_R^u$.}
	\label{fig5}
	\end{figure}

\begin{figure}[H]
	\centering
	\includegraphics[width=0.85\textwidth,height=0.4\textwidth]{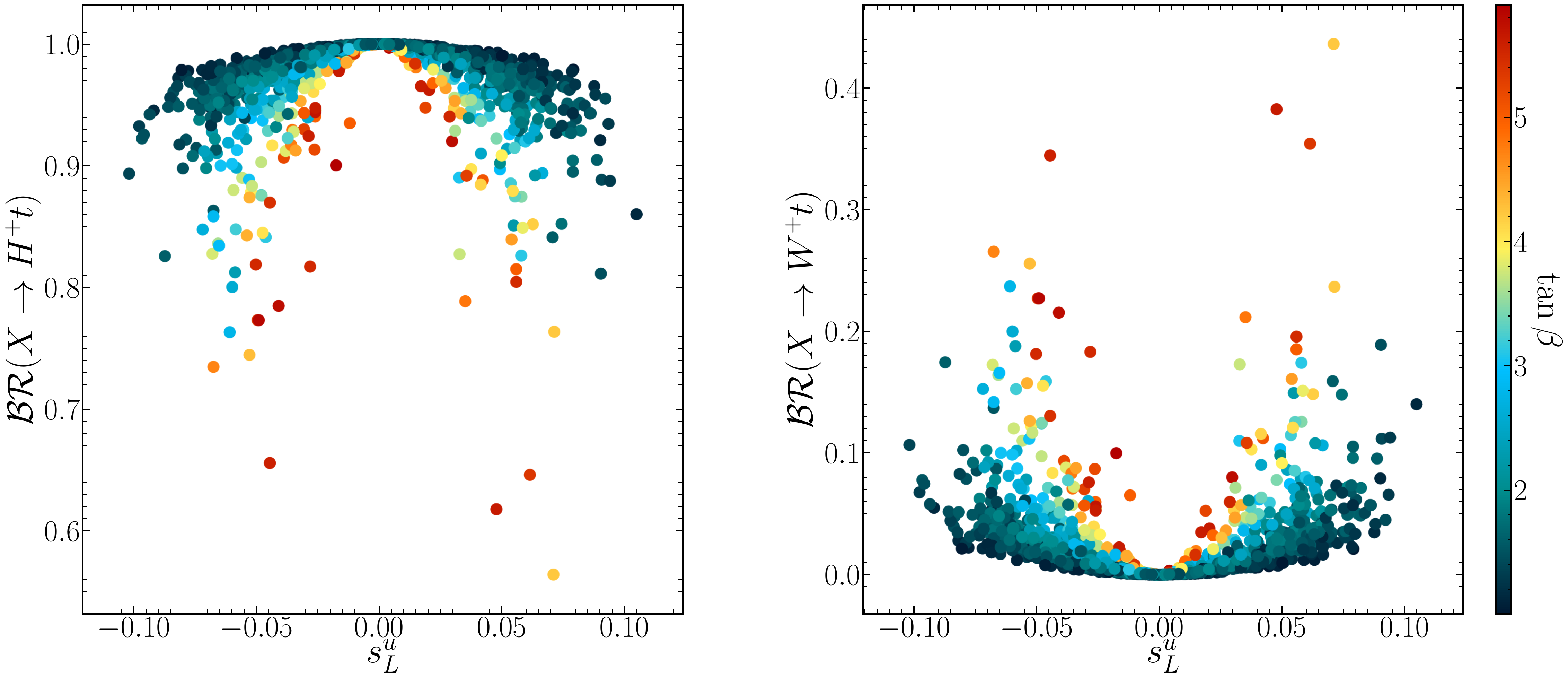}
	\caption{Similar to Fig.~\ref{fig5}, but with $\tan\beta$ shown in the colour bar.}
	\label{fig6}
	\end{figure}

	Again, we performed a comprehensive scan over the relevant 2HDM and VLQ parameters, as summarised in Tab.~\ref{table1}. In Figs.~\ref{fig5} and \ref{fig6}, we present  $\mathcal{BR}(X\to H^+t)$ (left) and $\mathcal{BR}(X\to W^+t)$ (right) as functions of $s_L^u$, with $s_R^u$ and $\tan\beta$ indicated by the colour bars, respectively. Unlike the doublet scenarios, the production of charged Higgs bosons from the VLQ $X$ in the triplet case can reach 100\% for various values of $s_L^u$. This is due to an enhancement in the left-handed coupling, which is proportional to $c_L^u$. Here, $c_L^u$ is close to one, as $s_L^u$ is constrained by EWPOs to remain small. Furthermore, the SM decay $\mathcal{BR}(X\to W^+t)$ can achieve a maximum of 44\%\footnote{For detailed expressions of the couplings, please refer to the Appendix.}.
	
	\subsection{2HDM with the $(TBY)$ Triplet}
	
	We now turn to the $(TBY)$ triplet scenario, which shares similarities with the $(T)$ singlet and $(TB)$ doublet cases within the 2HDM Type-II framework. The model is defined by specifying the new top quark mass and a mixing angle, $\theta_L^t$, with the remaining parameters computable from these inputs. Here, $\theta_R^t$ is determined using Eq.~(\ref{ec:rel-angle1}), and the mass of the VLQ $Y$  is given by \cite{Aguilar-Saavedra:2013qpa}:
	
	\begin{eqnarray}
	m_Y^2 &=& m_T^2 \cos^2\theta_L^t + m_t^2 \sin^2\theta_L^t \nonumber\\&=& m_B^2 \cos^2\theta_L^b + m_b^2 \sin^2\theta_L^b.
	\end{eqnarray}
	
	Using this relation between $m_T$ and $m_Y$, along with the mixing relations in Eq.~(\ref{TBY-mix}), the mass of the new bottom quark is derived as:
	
	\begin{eqnarray}
	m_B^2 &=& \frac{1}{8} \sin^2(2\theta_L^t)\frac{(m_T^2 - m_t^2)^2}{m_Y^2 - m_b^2} + m_Y^2.
	\end{eqnarray}
	
	With this, the down-type quark mixing angles, $\theta_L^d$ and $\theta_R^d$, are calculated using Eqs.~(\ref{ec:rel-angle1})--(\ref{TBY-mix}).
	
\begin{figure}[H]
	\centering
	\includegraphics[width=0.85\textwidth,height=0.4\textwidth]{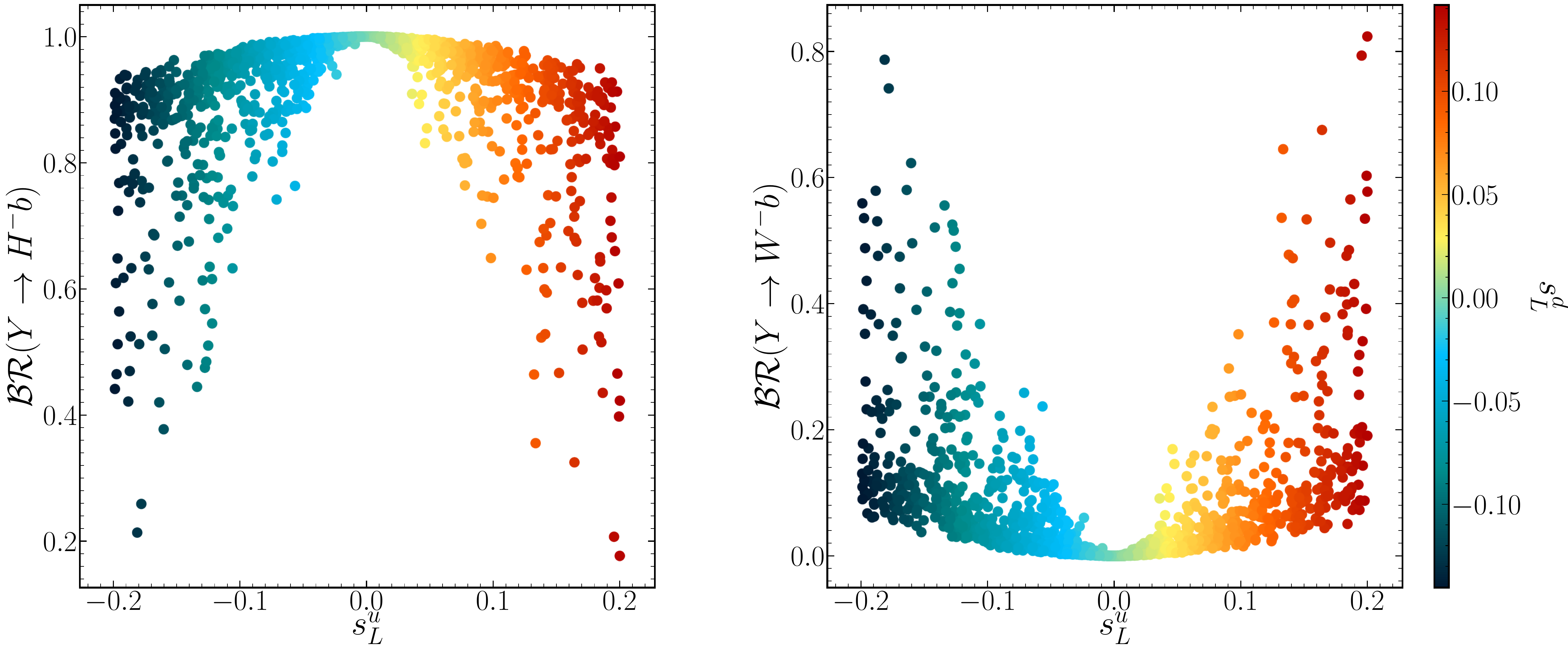}
	\caption{Scatter plots of parameter points that satisfy all imposed constraints in the $\mathcal{BR}(Y\to H^-b)$ versus $s_L^u$ (left) and $\mathcal{BR}(Y\to W^-b)$ versus $s_L^u$ (right) planes, with the colour bar indicating $s_L^d$.}
	\label{fig7}
	\end{figure}

\begin{figure}[H]
	\centering
	\includegraphics[width=0.85\textwidth,height=0.4\textwidth]{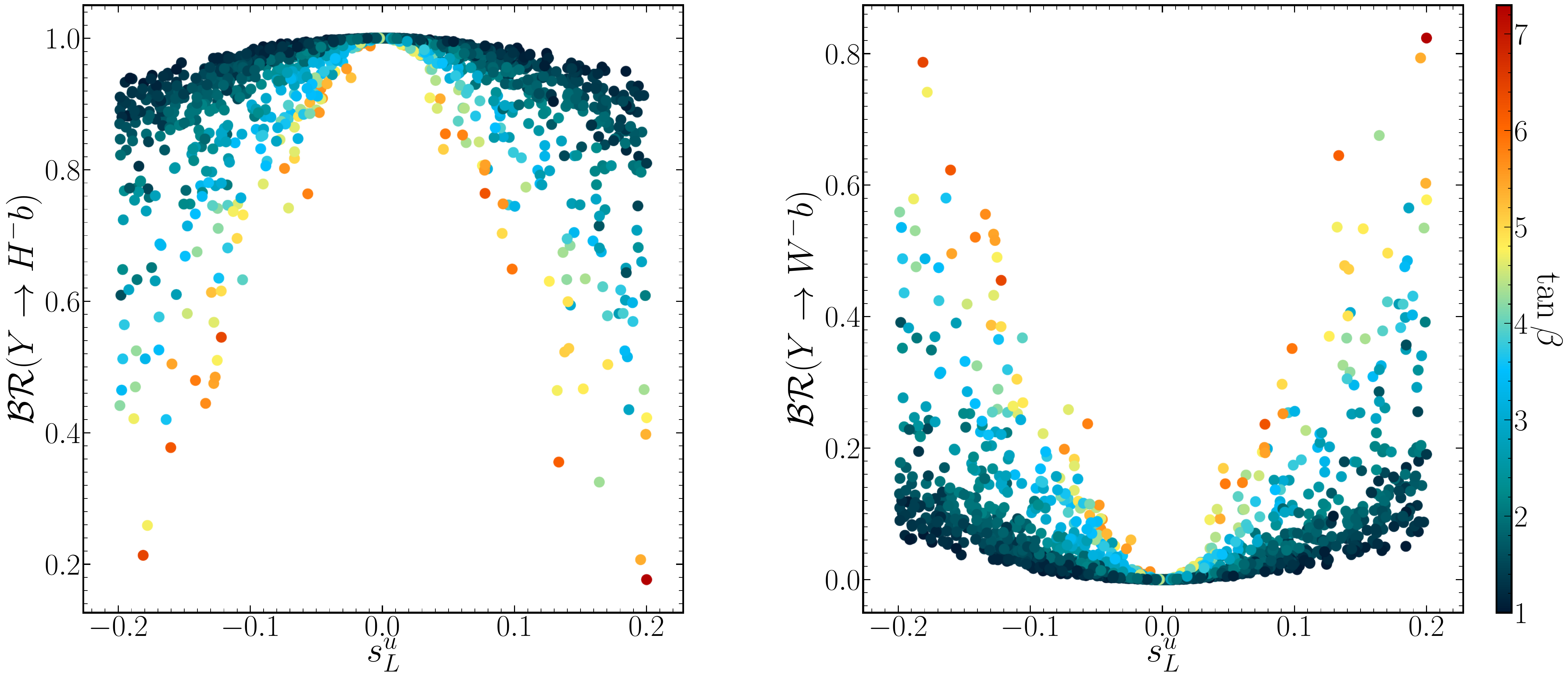}
	\caption{Similar to Fig.~\ref{fig7}, but with $\tan\beta$ shown in the colour bar.}
	\label{fig8}
	\end{figure}

	Figs.~\ref{fig7} and \ref{fig8} display $\mathcal{BR}(Y \to H^-b)$ (left) and $\mathcal{BR}(Y \to W^-b)$ (right) as functions of $s_L^u$, with $s_L^d$ and $\tan\beta$ indicated by the colour bars, respectively. Similar to the $(XTB)$ triplet, the production of charged Higgs bosons from VLQ-$Y$ in this triplet can also achieve a 100\%  ${\cal BR}$ for various $s_L^u$ values, owing to the enhancement of the left-handed coupling, which is close to unity due to the constraints on $s_L^u$ from EWPOs. In contrast, the SM decay channel $\mathcal{BR}(Y \to W^-b)$ can reach up to 88\%.
	
	\section{Discussion and Conclusion}
	
	Before concluding, we emphasise the distinctive signatures arising from the decays of the VLQs $X$ and $Y$ in both the SM extended with VLQs (SM+VLQ) and the 2HDM extended with VLQs (2HDM+VLQ). Our analysis highlights significant differences in collider phenomenology between these scenarios, primarily due to the additional decay modes introduced by the extended Higgs sector in the 2HDM.
	
	In the SM+VLQ scenario, pair production of $X$ quarks predominantly leads to the final state $X\bar{X} \to W^+ t\, W^- \bar{t}$, which, after the top quark decays $t \to W b$, results in $4W + 2b$-jets. Similarly, pair production of $Y$ quarks yields $Y\bar{Y} \to W^+ b\, W^- \bar{b}$, leading to $2W + 2b$-jets. These signatures involve standard processes with well-understood backgrounds.
	
	In contrast, the 2HDM+VLQ framework introduces additional decay possibilities due to the presence of the charged Higgs bosons $H^\pm$. The VLQs can decay into $H^\pm$ along with SM fermions, specifically, $X \to H^- t$ and $Y \to H^- b$. The charged Higgs bosons themselves exhibit rich decay patterns, notably $H^\pm \to t b$ with a ${\cal BR}$ reaching 100\%, or $H^\pm \to W^\pm A/H$ with significant probabilities. Here, $A$ and  $H$ predominantly decay into $t\bar{t}$ due to their substantial masses.
	
	These new decay modes lead to novel and complex final states. For example, in $X\bar{X}$ pair production, the process $X\bar{X} \to H^+ t\, H^- \bar{t}$ followed by $H^\pm \to t b$ and subsequent top quark decays results in a final state of $4W + 6b$-jets. Alternatively, if $H^\pm$ decays into $W^\pm A/H$ with $A/H \to t\bar{t}$, the final state includes up to $8W + 6b$-jets. For $Y\bar{Y}$ production, similar processes yield final states with $2W + 6b$-jets or $6W + 6b$-jets, depending on the decay pathways of the $H^\pm$ states.
	
	The increased number of $W$ bosons in these final states has significant implications for collider signatures. Although the probability of all $W$ bosons decaying leptonically is relatively low given that the leptonic $\mathcal{BR}$s of the $W$ boson are approximately 10.7\% for decays into electrons and 10.6\% for decays into muons, the resulting events are highly distinctive. For instance, scenarios where two same-sign $W$ bosons decay leptonically into electrons or muons produce same-sign dilepton events accompanied by light-quark jets and multiple $b$-jets. These events are particularly significant because the SM background for such processes is low, making them excellent channels for probing new physics. Even if only a subset of the $W$ bosons decay leptonically, the presence of multiple leptons and $b$-jets provides unique signatures that can be effectively utilised in experimental analyses.
	
	Our findings underscore the importance of considering these non-standard decay channels in VLQ searches at the LHC. The interplay between the VLQs and the extended Higgs sector significantly impacts the decay patterns and experimental signatures, necessitating careful consideration in experimental studies. Constraints from EWPOs play a crucial role in shaping the decay profiles of the VLQs, influencing both their mass spectra and couplings.
	
	In summary, our 2HDM+VLQ framework presents a rich phenomenology distinct from the SM+VLQ scenario. The differences in final states characterised by an increased number of $W$ bosons and $b$-jet combined with the possibility of same-sign leptons offer compelling signatures for experimental investigation. Future collider experiments, with enhanced detection capabilities and advanced analysis techniques, will be instrumental in probing these scenarios and potentially uncovering new physics beyond the SM of the kind advocated here.

	\section{Acknowledgments}
	SM is supported in part through the NExT Institute and the STFC Consolidated Grant ST/L000296/1.
	\appendix
	\section{Appendix}
	\subsection{Decay Patterns of the Extended Higgs Bosons}
	\begin{figure}[H]
	\centering
	\includegraphics[width=\textwidth, height=0.35\textwidth]{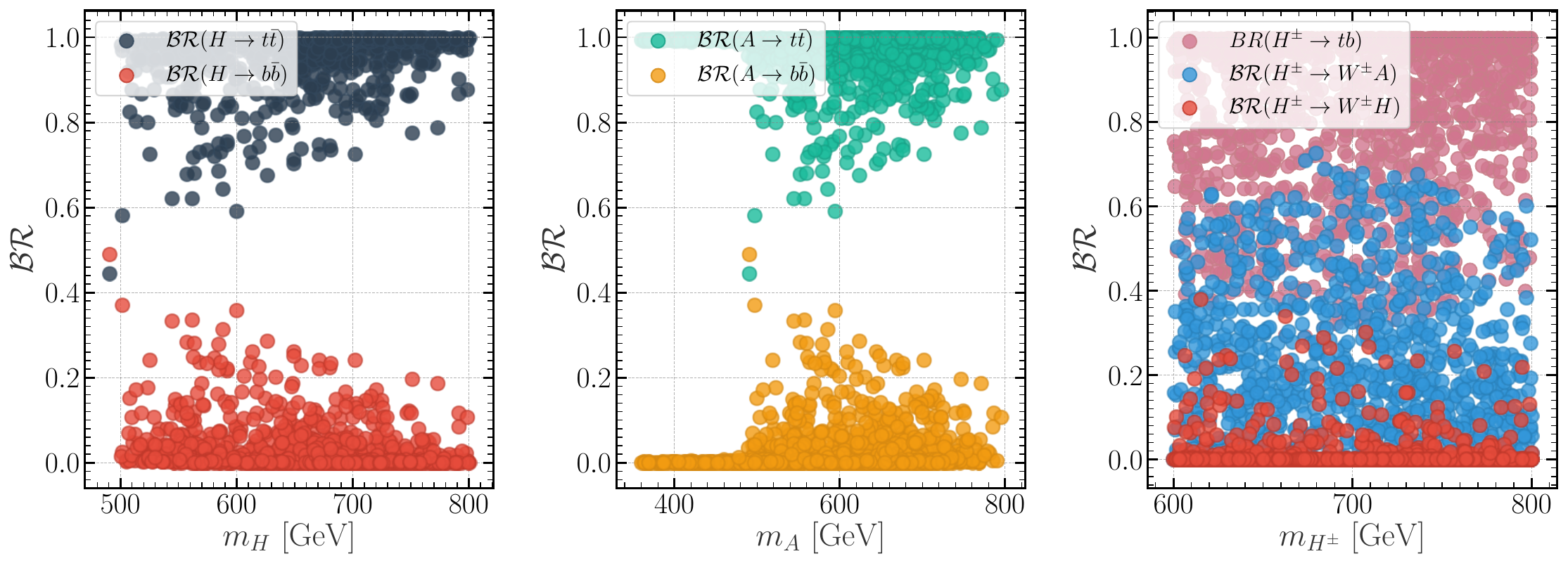}
	\caption{$\mathcal{BR}$s of $H$, $A$, and $H^\pm$ for the allowed parameter points discussed in the main text. The left panel shows $\mathcal{BR}(H \to tt, bb)$ as a function of $m_H$, the middle panel shows $\mathcal{BR}(A \to tt, bb)$ as a function of $m_A$, and the right panel shows $\mathcal{BR}(H^\pm \to tb, W^\pm A, W^\pm H)$ as a function of $m_{H^\pm}$.}
	\label{fig_app}
	\end{figure}
	In this appendix, we provide a detailed discussion of the decay patterns of the additional Higgs bosons in the model, focusing on the charged Higgs boson $H^\pm$ and the neutral Higgs bosons $H$ and $A$. These decay channels are crucial for understanding the phenomenology of the VLQs $X$ and $Y$, as they significantly impact the final states observable at collider experiments.
	
	Fig.~\ref{fig_app} illustrates the $\mathcal{BR}$s  of $H^\pm$, $H$, and $A$ for the allowed parameter points discussed in the main text. The masses of $H$ and $A$ vary in the range of $400$ to $800~\text{GeV}$, while the mass of $H^\pm$ spans from $600$ to $800~\text{GeV}$.
	
	As depicted in the figure, the charged Higgs boson $H^\pm$ can decay into $tb$ with a $\mathcal{BR}$ reaching $100\%$. Additionally, $H^\pm$ can decay into $W^\pm A$ and $W^\pm H$ with maximum $\mathcal{BR}$s of approximately $70\%$ and $38\%$, respectively. These decay modes are important because they open up new channels for the production of heavy Higgs bosons.
	
	As expected for heavy Higgs bosons, both $H$ and $A$ decay predominantly into top-antitop quark pairs ($t\bar{t}$), with $\mathcal{BR}$s approaching $100\%$. In certain regions of the parameter space, the decay into $b\bar{b}$ can also have a substantial $\mathcal{BR}$, reaching approximately $50\%$.

	\renewcommand{\thetable}{\Roman{table}} 
	\subsection{Lagrangian in the mass basis}
	
	As mentioned, after EWSB, we are left with five Higgs bosons: two-CP even ones, $h$ and $H$, one CP-odd one, $A$, and then a pair of charged Higgs states, $H^\pm$. We now collect the Lagrangian in the  mass basis in the general 2HDM Type-II supplememented by VLQs.

	\subsection*{Light-heavy interactions}
	
	Here is the relevant Lagrangian for such interactions:
	\begin{eqnarray}
	\mathcal{L}_W & = & - \frac{g}{\sqrt{2}} \overline{Q} \gamma^\mu (V^L_{Qq}P_L + V^R_{Qq}P_R )q W^+_\mu \nonumber\\
	&& - \frac{g}{\sqrt{2}} \overline{q} \gamma^\mu (V^L_{qQ}P_L + V^R_{qQ}P_R )Q W^+_\mu   + H.c.\nonumber \\
	\mathcal{L}_{H^+} &=& - \frac{g m_Y}{\sqrt{2}M_W}\overline{Y} (\cot\beta Z^L_{bY} P_L + \tan\beta Z^R_{bY} P_R ) b H^+\nonumber\\
	&&  - \frac{g m_X}{\sqrt{2}M_W}\overline{t} (\cot\beta Z^L_{Xt} P_L + \tan\beta Z^R_{Xt} P_R ) X H^+ \nonumber\\
	&&+h.c.
	\end{eqnarray}
	where the relevant couplings are given in Tabs.~II--V.

	\begin{eqnarray}
	\begin{array}{c|cccc}
	&	V_{Xt}^L &	V_{Xt}^R     \\ \hline
	
	(XT) & -s_L e^{-i\phi} &-s_R e^{-i\phi}	 \\
	
	(XTB) & - \sqrt{2} s_L^u e^{-i\phi} & - \sqrt{2} s_R^u e^{-i\phi} \\
	
	\end{array} \nonumber 
	\end{eqnarray}\captionof{table}{Heavy-light couplings to the $W^\pm$ boson.}	
	
	\begin{eqnarray}
	\begin{array}{c|cccc}
	&	V_{bY}^L &V_{bY}^R  \\ \hline 
	
	(BY)  & -s_L e^{i\phi} &-s_R e^{i\phi}  \\
	(TBY)  &  -\sqrt{2}s_L^d  e^{i\phi} & -\sqrt{2}s_R^d  e^{i\phi}
	\end{array} \nonumber
	\end{eqnarray}\captionof{table}{Light-heavy couplings to the $W^\pm$ boson.}

	\begin{eqnarray}
	\begin{array}{c|cc}
	& Z^L_{Xt} & Z^R_{Xt} \\ \hline 
	(XT)&  s_R e^{-i\phi}   &0    \\ 
	(XTB)&   c_L^u & 0 
	\end{array}  
	\nonumber
	\end{eqnarray}	
	\captionof{table}{Heavy-light couplings to the $H^\pm$ boson.}
	
	\begin{eqnarray}
	\begin{array}{c|cc}
	& Z^L_{bY} & Z^R_{bY}\\ \hline 
	(BY)&  s_R e^{-i\phi}   &0  \\ 
	
	(TBY)&   c_L^u & 0 
	\end{array} 
	\nonumber
	\end{eqnarray}	
	\captionof{table}{Light-heavy couplings to the $H^\pm$ boson.}
	\newpage
	\bibliography{main}
	\bibliographystyle{JHEP}
	\end{document}